\documentclass{jaa}

\usepackage{graphicx}
\usepackage{xcolor}

\begin{document}\sloppy

\title{Optical detection of a GMRT-detected candidate high-redshift radio galaxy with 3.6-m Devasthal optical telescope}


\author{A. Omar\textsuperscript{1}, A. Saxena\textsuperscript{2}, K. Chand\textsuperscript{1}, A. Paswan\textsuperscript{1,3}, H. J. A. R\"ottgering\textsuperscript{2}, K. J. Duncan\textsuperscript{2}, T. S. Kumar\textsuperscript{1}, B. Krishnareddy\textsuperscript{1}, 
J. Pant\textsuperscript{1}}
\affilOne{\textsuperscript{1}Aryabhatta Research Institute of observational-sciences, Manora Peak, Nainital 263001, India. \\}
\affilTwo{\textsuperscript{2}Leiden Observatory, Leiden University, PO Box 9513, NL-2300 RA Leiden, The Netherlands.\\}
\affilThree{\textsuperscript{3}Inter-University Centre for Astronomy and Astrophysics, Post Bag 4, Ganeshkhind, Pune 411007, India. \\}

\onecolumn{

\maketitle

\corres{aomar@aries.res.in}

\msinfo{December 2018}{February 2019}

\begin{abstract}
We report optical observations of TGSS J1054+5832, a candidate high-redshift ($z=4.8\pm2$) steep-spectrum radio galaxy, in $r$ and $i$ bands using the faint object spectrograph and camera mounted on 3.6-m Devasthal Optical Telescope (DOT). The source previously detected at 150 MHz from Giant Meterwave Radio Telescope and at 1420 MHz from Very Large Array has a known counterpart in near-infrared bands with $K$-band magnitude of AB 22. The source is detected in $i$-band with AB$24.3\pm0.2$ magnitude in the DOT images presented here. The source remains undetected in the $r$-band image at a 2.5$\sigma$ depth of AB 24.4 mag over an $1.2''\times1.2''$ aperture. An upper limit to $i-K$ color is estimated to be $\sim$2.3, suggesting youthfulness of the galaxy with active star formation. These observations highlight the importance and potential of the 3.6-m DOT for detections of faint  galaxies. 
\end{abstract}


}


\doinum{12.3456/s78910-011-012-3}
\artcitid{\#\#\#\#}
\volnum{000}
\year{0000}
\pgrange{1--}
\setcounter{page}{1}
\lp{1}

\section{Introduction}

The ultra steep spectrum (USS) sources at radio wavebands are often associated with high redshift radio galaxies (HzRGs; e.g., R\"ottgering {\em et al.} 1994; De Breuck {\em et al.} 2000; Saxena {\em et al.} 2018a). The USS radio sources are characterized by a radio spectrum with spectral index $\alpha < -1.3$ ($S_{\nu} \propto \nu^{\alpha}$; where $S$ is flux density and $\nu$ is radio frequency). The spectral index of HzRGs normally steepens with increasing redshift (e.g., Ker {\em et al.} 2012) resulting in to a $\alpha-z$ relationship, which can be used to identify candidate HzRGs from large radio surveys. At high redshifts, the rest-frame ultraviolet Lyman break (between 91.2 - 121.6 nm) due to absorption of ultraviolet photons by Hydrogen gas in the interstellar medium of a galaxy shifts to visible (optical) bands and serves as a powerful technique for detecting high-$z$ galaxies and obtaining photometric redshifts (see Steidel {\em et al.} 1996; Brammer {\em et al.} 2008 for examples). The Balmer break near 400 nm (rest-frame), where optical flux from galaxies drops although the break is less strong compared to the Lyman break, can also be used to infer high-$z$ nature (e.g., Franx et al. 2003). The radio galaxies which are seen to be dominated by an old stellar population (Best {\em et al.} 1998; Rocca-Volmerange {\em et al.} 2004; Seymour {\em et al.} 2007), show a remarkably tight $K-z$ relation out to high redshifts in the sense that high-$z$ galaxies become fainter in near-infrared bands (e.g., Lilly \& Longair 1984; Jarvis {\em et al.} 2001a; Willott {\em et al.} 2003; Saxena {\em et al.} 2018b). Additionally, relationships between optical-NIR colors and $z$ (e.g., De Breuck {\em et al.} 2002, Schmidt {\em et al.} 2006) are also seen for HzRGs. Therefore, NIR faintness and color selection in addition to ultra-steep spectrum selection at radio are routinely used to identify HzRGs candidates.

HzRGs trace the most massive and the brightest galaxies in early epochs of large-scale cosmological structure formation (e.g., Venemans {\em et al.} 2003, Overzier {\em et al.} 2009). A recent review on HzRGs and their environments is provided by Miley \& De Breuck (2008). The density of radio galaxies falls off rapidly beyond redshift 3 (Dunlop \& Peacock 1990; Jarvis {\em et al.} 2001b; Willott {\em et al.} 2001; Rigby {\em et al.} 2011, 2015). Observationally, not much is known about the physical properties such as star formation rates, black hole masses and nuclear activities of $z>4$ radio galaxies, largely due to very small number statistics. The lower redshift counterparts at $z<4$ are often found in a cluster-like environment, rich in dust and gas, and with high star formation rates (e.g., Best {\em et al.} 1998; Pentericci {\em et al.} 2000; Carilli {\em et al.} 2002; R\"ottgering {\em et al.} 2003;  Orsi {\em et al.} 2016). The high radio luminosities of HzRGs are powered by the central supermassive black holes in the active galactic nuclei hosted by these galaxies. As HzRGs are considered to be the progenitors of the most massive elliptical galaxies that are observed in the local Universe, their detections are significant for theories of galaxy and large structure formation and subsequent evolution. 

Recently, Saxena {\em et al.} (2018a) identified candidate HzRGs based on the detections in the TIFR GMRT Sky Survey First Alternative Data Release (TGSS ADR1; Intema {\em et al.} 2017) images made at 150 MHz using the Giant Meterwave Radio Telescope (GMRT), 370 MHz images made using the Very Large Array (VLA), and the Faint Images of the Radio Sky at Twenty centimetres (FIRST; White {\em et al.} 1997) images near 1400 MHz made using the VLA. One of the sources, TGSS J1054+5832 ($R.A.=10^{h}54^{m}29.5^{s};~Dec.=+58^{d}32^{m}27.6^{s} ~J2000$), in that sample has a spectral index of $\alpha^{150}_{370} = -1.3$ and $\alpha^{370}_{1400} = -1.5$. The source has a flux density of $100$ mJy at 150 MHz, $30$ mJy at 370 MHz and $4$ mJy at 1400 MHz. This source is also identified as 7C 1051+5848. TGSS J1054+5832 lies in the well-studied Lockman Hole field and was initially reported as a USS radio source by Mahony {\em et al.} (2016). This source has detections in UKIDSS Deep Extragalactic Survey (DXS; Dye {\em et al.} 2006) $J$ and $K$ bands images and in the {\em Spitzer} IRAC bands at 3.6 $~\mu$m and 4.5 $\mu$m wavelength. The $K$-band magnitude is estimated to be AB$\sim$22. Majority of the known HzRGs have their K-band magnitude brighter than 20 mag. Only the two known HzRGs with magnitudes near $K=22$  have $z\sim$5  (van Breugel {\em et al.} 1999, Saxena {\em et al.} 2018b). This makes J1054+5832 a $K$-band faint radio galaxy, very likely to lie at a redshift similar to the other two known sources near or beyond $z=5$. Following the $K-z$ relation, TGSS J1054+5832 is predicted to be at high redshift ($z>4$). Further, TGSS J1054+5832 has no known detection in the publicly available optical bands. It is not detected in PanSTARRS1 images (Chambers {\em et al.} 2016) down to AB 22.3  in $z$-band and AB 23.1 in $i$-band. A photometric redshift of $z=4.8\pm2.0$ is derived for this object, based on the upper limits in optical bands and detections in the NIR bands (Saxena {\em et al.} 2018a). The optical detections of distant radio galaxies are challenging and require moderate and sometimes the largest aperture telescopes available presently. 

In this work, we report observations of TGSS J1054+5832 using the 3.6-m Devasthal optical telescope (DOT), commissioned in 2016. These observations were taken in Sloan Digital Sky Survey (SDSS) standard $r$ and $i$ band filters. The layout of this paper is as follows. We describe the instrument and the observational setup in Section 2. The results of the observations and preliminary discussions are presented in Section 3. In Section 4, we present a summary of our findings.  

\section{Instrument description and observations}

\begin{table*}[htb]
\caption{Expected peak efficiency of the instrument system in different bands}\label{tableExample} 
\begin{tabular}{lcccccc}
\topline
Band (Wavelengths) & $T_{\mathrm{atmosphere}}$ & $R_{\mathrm{telescope}}$ & $T_{\mathrm{ADFOSC}}$ & $T^{peak}_{\mathrm{filter}}$ & $QE_{\mathrm{CCD}}$ & $T^{peak}_{\mathrm{system}}$ \\
(nm) & (\%) & (\%) & (\%) & (\%) & (\%) & (\%) 
\\\midline
u ($360\pm28$) & 52 & 67 & 84 & 82 & 40 & 9.6\\
g ($464\pm64$) & 76 & 69 & 84 & 90 & 85 & 33.7\\
r ($612\pm58$) & 83 & 69 & 86 & 95 & 90 & 42.1\\
i ($744\pm61$) & 88 & 70 & 87 & 94 & 90 & 45.3\\
z ($890\pm61$) & 91 & 65 & 80 & 95 & 60 & 27.0\\
z (1000) & 91 & 64 & 68 & 95 & 10 & 3.8\\
\hline
\end{tabular}
\tablenotes{Notes: (1) Wavelengths for the first 5 rows are given as the center wavelength$\pm$half bandwidth of the filter pass-bands. The last row is for 1000 nm wavelength falling in the pass-band of the $z$-filter. (2) $T_{\mathrm{atmosphere}}$ for Devasthal are taken from Pandey {\em et al.} (2018). As the atmospheric transmission decreases sharply below 400 nm, the peak efficiency indicated for center of the $u$-band is uncertain. (3) $R_{\mathrm{telescope}}$ are based on the measured reflectivity of the freshly coated primary mirror and assuming that the secondary mirror has 80\% reflectivity in all bands. (4) $T_{\mathrm{ADFOSC}}$ in $u$, $g$, $r$, and $i$ bands are based on the measurements on the coated samples and that in the $z$-band is extrapolated value from the transmission curve measured at shorter wavelengths. (5) $T_{\mathrm{filter}}$ and $QE_{\mathrm{CCD}}$ are based on the measurements on the actual supplied filters and the actual supplied CCD sensor by their respective vendors.}
\end{table*}

\begin{figure*}
\begin{center}
       \includegraphics[scale=0.5]{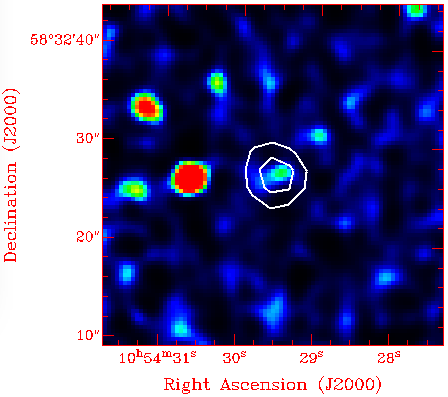}
        \includegraphics[scale=0.5]{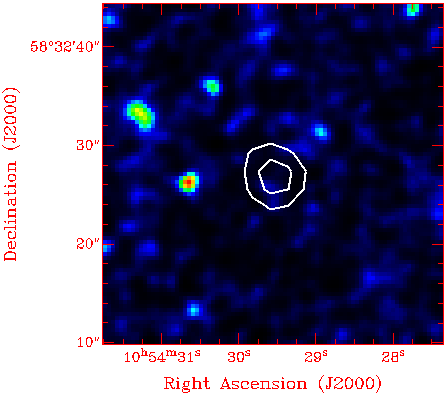}
        \caption{The DOT $i$-band image (left) and $r$-band image (right) of TGSS J1054+5832 with 1.4 GHz FIRST radio contours (1 and 2 mJy beam$^{-1}$) overlay. The galaxy is detected in $i$-band but not detected in $r$-band.}
\end{center}       
\end{figure*}

The observations presented here were made using the Faint Object Spectrograph and Camera (FOSC) type back-end instrument on the 3.6-m Devasthal optical telescope (DOT; $29.36^{0}$N, $79.69^{0}$E, 2426-m above mean sea level) to test imaging capabilities of the instrument system. DOT with the 3.6-m aperture is the largest optical telescope in India. It was recently commissioned for science observations. The technical details and scientific capabilities of the telescope are presented elsewhere (Ninane {\em et al.} 2012; Omar {\em et al.} 2017; Kumar {\em et al.} 2018). The primary mirror made of Zerodur is periodically coated with Aluminium at Devasthal using the in-country developed coating plant (Bheemireddy et al. 2016). The secondary mirror made of Sitall was coated with Aluminium and Silicon-dioxide protective coating in 2010 by the supplier and the measured average reflectivity at that time was 87\% in the wavelength range of 400 - 1000 nm. The present reflectivity of the secondary mirror is not known but expected to have degraded nominally. The telescope has measured rms tracking accuracies in open-loop between $0.16''$ and $0.42''$ over 15 minutes of observations at different elevations and atmospheric conditions. The best images obtained from the telescope have Full-Width Half-Maxima (FWHM) down to $0.4''$. The telescope is expected to make routine detections of faint celestial sources down to 25 mag in $r$ and $i$ bands in a reasonable integration time. 

The FOSC-type instrument (hereafter ADFOSC; ARIES-Devasthal FOSC), developed in ARIES (Aryabhatta Research Institute of observational sciences), Nainital, is presently undergoing commissioning. Details of the ADFOSC and the Charge-Coupled Device (CCD) camera are presented elsewhere (Omar {\em et al.} 2012; Omar {\em et al.} 2019). The ADFOSC has a collimator and focal-reducer optics, converting the f/9 beam from the DOT to $\sim$f/4.3 beam. The optical path of the ADFOSC has only transmission optics. It can be used for imaging and spectroscopy of faint celestial sources in the wavelength range 350-1050 nm. Various filters, slits and grisms can be inserted in the path. The optical design of the spectrograph is optimized to obtain 80\% encircled energy within $0.4''$ diameter at the image plane, anywhere within the central $10'$ diameter.  The ADFOSC has a $4096\times4096$ deep-depletion Charge-Coupled Device (CCD) camera, optimized for observations towards the red bands ($r,i,z$). The CCD chip ($version$ 231-84) is procured from E2V Inc. U.K., and has broadband coating optimized for fringe suppression in the red bands. With a plate-scale of $13.3''$~mm$^{-1}$ at the ADFOSC focal-plane, each pixel (15 $\mu$m) on the CCD corresponds to $\sim0.2''$. The expected peak efficiency of the instrument system ($T^{peak}_{\mathrm{system}}$) estimated as the product of atmospheric transmission at Devasthal ($T_{\mathrm{atmosphere}}$), telescope reflectivity ($R_{\mathrm{telescope}}$), transmission of the ADFOSC optics ($T_{\mathrm{ADFOSC}}$), peak transmission of the SDSS filters ($T^{peak}_{\mathrm{filter}}$), and the CCD quantum efficiency ($QE_{\mathrm{CCD}}$) are given in Table~1. The efficiencies at the atmospheric cut-off of the visible light near 360 nm and the CCD response cut-off near 1000 nm are expected at $\sim$10\% and $\sim$4\%, respectively. The instrument system is expected to have reasonably good average efficiency of $\sim$20\% between 800-1000 nm (part of $z$-band). The ADFOSC is therefore expected to be a sensitive instrument that can be employed to obtain optical detections of high-redshift objects, which are generally brighter towards the redder bands compared to the bluer bands. 

The DOT observations were carried out in the dark (new moon) night of April 16, 2018. The relative humidity was less than 60\% and the night-sky was clear and photometric during the observations. Multiple frames of the target field were observed. As the telescope auto-guider was not used, integration time in each frame was kept to 300-sec. A total of 4 frames in $r$-band and 12 frames in $i$-band were observed. No star-trailing was noticed in the frames. The observations were constrained by the spare time available between some planned tests with the ADFOSC and therefore were not optimized. As a consequence, standard photometric standard fields could not be observed. The source was observed at zenith distance between $30^{\circ}  - 40^{\circ} $ after the transit. The CCD camera controller was operated with near-unity gain to obtain accurate photometry of extremely faint celestial sources and with low read-out frequency ($\sim$160 kHz) to keep read-out noise below 10 $e^{-}$ rms. The CCD is cooled to $-120^{\circ}$ C using a closed-cycle cryo-cooling thermal engine to have negligible dark noise below 0.001 $e^{-}$ per second. The on-chip binning of $2\times2$ pixels was used during the observations.  

Standard procedures of CCD data reduction were performed that included the following steps. All the frames were corrected for the CCD bias level. The master flat in $r$-band was made using the twilight sky frames, while that in $i$-band was made using the median combination of the sky frames obtained at different nights. No fringing was noticed in the $i$-band image. Based on our experience with the data reduction with ADFOSC, night-sky flat fielding gives better results than twilight flat fielding in redder bands. The frames were co-aligned and median combined. The median combination effectively filters out spurious signals such as cosmic-ray hits. The images were inverted along horizontal axis and rotated using DS9 software in order to bring north to the top and east to the left. The images were registered for the equatorial coordinate system using the SDSS images available for the same region. The digital counts in the final images were converted to the flux unit (Jy) and the AB magnitude system using the common stellar-like objects in the calibrated SDSS images and the DOT images. Total of nine comparison objects between 20-22 mag were used to obtain photometric calibration. The photometric calibration (Jy per digital count per sec) is taken as the median and the error is taken as the dispersion of the distribution of the values obtained from the nine comparison objects. The estimated magnitudes from the DOT images are therefore on the AB magnitude system. We estimated that 1 digital count per sec corresponds to $1.04\pm0.04~\mu$Jy in the $i$-band and $1.16\pm0.05~\mu$Jy in the $r$-band, indicating that the $i$-band has marginally higher efficiency compared to the $r$-band.  This efficiency trend is broadly consistent with the expected trend inferred from Table~1. We also estimated the night sky brightness as $\sim$20.96 mag arcsec$^{-2}$ ($\sim$15~$\mu$Jy arcsec$^{-2}$) in the $r$-band and $\sim$20.43 mag arcsec$^{-2}$ ($\sim$24~$\mu$Jy arcsec$^{-2}$) in the $i$-band using the average background counts in the DOT images. For comparison, SDSS 25-percentile average sky brightness values are 21.04 mag arcsec$^{-2}$ in the $r$-band and $20.36$ mag arcsec$^{-2}$ in the $i$-band. It implies that the night sky at Devasthal is dark at par with the Apache Point Observatory site of the SDSS telescope.

While comparing the NIR images and the DOT images, we noticed that several common detections in the field at the faintest levels are made at $2.5\sigma$ level in 3x3 pixel (1 pixel = $0.4''$) aperture in the DOT images. This corresponds to the faintest possible detections at AB$\sim$24.4 mag or $\sim$0.7 $\mu$Jy in the $r$-band and at AB$\sim$24.9 mag or $\sim$0.4 $\mu$Jy in the $i$-band. The FWHM in the stacked images is $\sim$1.5'' in both $r$ and $i$ band images.  

\section{Results and discussions}

\begin{figure}
\begin{center}
       \includegraphics[scale=0.52]{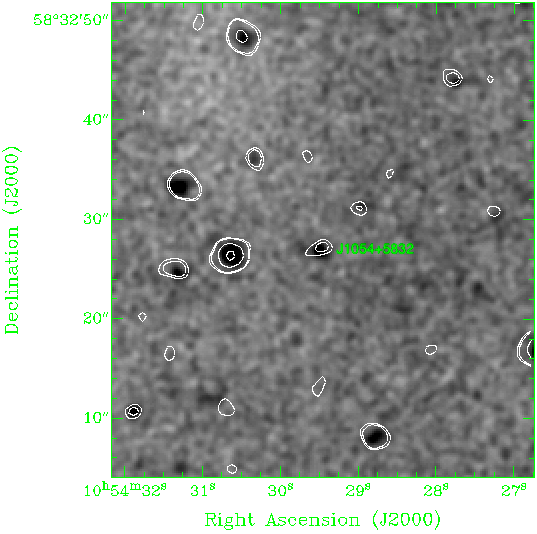}
        \caption{The DOT $i$-band contours overlaid upon the UKIDSS DXS $J$-band image. The first contour is at 3$\sigma$ level and the rests are arbitrarily chosen to show other sources of different intensities. The HzRG near the center is marked as J1054+5832.}
\end{center}       
\end{figure}

The images obtained from the DOT in $r$ and $i$ bands were smoothed by the Gaussian kernel of 3x3 pixels ($1.2''\times1.2''$) to enhance probability of detecting weak emission. The smoothed images are presented in Fig.~1. A point-like source is detected in the $i$-band, matching with the location of the radio source TGSS J1054+5832. We associate this emission with the candidate HzRG TGSS J1054+5832. No clear detection is made at this location in the $r$-band image. The NIR detections in the $J$ and $K$ bands were already identified at the location of the radio source in Saxena {\em et al.} (2018a). A contour overlay of the DOT $i$-band image on the DXS $J$-band is presented in Fig.~2. The peak in the $i$-band detection matches with that in the $J$-band detection within $0.4''$. It is also interesting to see that both $i$ and $J$ band images show a hint of extended emission towards the east. The flux densities and AB magnitudes of the candidate HzRG are estimated in $i$-band using aperture photometry. The $i$-band flux for the object identified with the $J$-band counterpart of TGSS J1054+5832 is estimated to be $0.70\pm0.15~\mu$Jy or AB$24.3\pm0.2$ mag within an aperture of $1.5''$ radius. Although, the detection reported here in the $i$-band for TGSS J1054+5832 is marginal, it is a statistically significant detection given that the locations of the radio and previously reported NIR detections matches with that of the optical detection reported here. The Galactic extinction is negligible in this direction ($\sim$0.014 mag in $r$ band and $\sim$0.010 mag in $i$ band; Schlafly {\em et al.} 2011) compared to the errors in the optical flux estimates for the radio source made here. Therefore, an extinction correction was not necessary and was not applied. 

A SED model fit predicted the redshift for this source as $4.8\pm2$ based on the NIR detections and the limits in the $i,z,y$ bands (Saxena {\em et al.} 2018a). The flux in the $J$ band was estimated to be $\sim$3~$\mu$Jy. Therefore, there is a significant drop in flux between $J$-band and $i$-band. A non-detection in $r$-band indicates that flux drop is continuing towards the shorter wavelengths. This flux drop may be either due to the Lyman-break occurring between $J$-band and $i$-band or an intrinsic SED of a red galaxy such as a passively evolving massive elliptical or a post star-burst galaxy. With the limited information, it is not possible at the moment to distinguish between the two scenarios. The former scenario will put this galaxy to a high redshift ($z\geq5$). A significantly refined estimate for the photometric redshift is also not viable at the moment in absence of a flux estimate between $J$ and $i$ band (e.g., $z$ band) and a better limit in $r$-band. A spectroscopic determination of the redshift is also highly desirable. 

The $i-K$ color of TGSS J1054+5832 is estimated to be $\sim$2.3. This color is not corrected for intrinsic and extrinsic dust extinction and therefore represents an upper limit. It is known from the work of Lilly \& Longair (1984) that radio galaxies become bluer at high redshifts beyond 1 compared to their lower redshift counterparts.  De Breuck {\em et al.} (2002) found that the observed $R-K$ colors of the known radio galaxies become relatively bluer near their formation epochs at high redshifts, broadly consistent with the predictions from the evolution models. The color evolution of high-redshift galaxies is also modelled in Schmidt {\em et al.} (2006) for various formation scenarios at different redshifts. In the rest-frame, galaxies are predicted to be bluer and brighter in optical bands both at low redshifts $z<1$ and at high redshifts near their formation epochs $z>4$. The $i-K$ colors are predicted between 2 and 3 near their formation epochs and as the galaxy evolves, the $i-K$ color becomes redder (i.e., between 5 and 7) at redshifts between 1 and 3. At lower redshifts below 1, the $i-K$ colors are again predicted to be bluer (i.e. below 4) due to more recent episodes of star formation. The bluer and brighter appearance in the optical bands at high redshift near their formation epochs is due to initial burst of star formation.  It is worth to point out that at $z\sim4$, the $i$ and $K$ bands sample the rest-frame far ultraviolet (FUV) and $g$-band light respectively. The local galaxies with young star formation are bright in FUV and display strong blue FUV-optical colors (e.g., Ree {\em et al.} 2011). A relatively bluer $i-K$ color for TGSS J1054+5832 suggests youthfulness of this galaxy with active star formation, perhaps at high redshift.

Finally, slightly extended nature of the source as indicated by both DXS $J$-band and DOT $i$-band images could be an indication of a disturbed morphology created in tidal interaction with other galaxies in the vicinity. The relatively blue color of this object also favours recent star formation which could be triggered by tidal interaction. Such a scenario also favours high radio luminosity of the object linked to the triggering of intense active galactic nuclei related activities. The morphology of this object remains to be verified via deeper and higher angular resolution optical observations.

\section{Summary and concluding remarks}

A statistically significant optical detection in SDSS $i$-band was found for the steep spectrum candidate HzRG TGSS J1054+5832 using the ADFOSC instrument on the 3.6-m DOT. The $i$-band AB magnitude for the radio source is estimated to be $24.3\pm0.2$. The known correlations for HzRGs and a preliminary analysis already indicate that TGSS J1054+5832 is a highly probable $z\geq4$ radio galaxy. With the $K$-band magnitude of 22, TGSS J1054+5832 is a potential HzRG similar to the two known $K$-band faintest HzRGs near or beyond $z=5$. The $i-K$ color of TGSS J1054+5832  is estimated to be $\sim$2.3, which suggests youthfulness of this object. TGSS J1054+5832 is not yet a spectroscopically confirmed HzRG and therefore a spectroscopic confirmation for the redshift is highly desirable. Further deeper photometric and spectroscopic observations of this source are being carried out to understand the nature of this source.  

The optical detection of the optically-faint radio source TGSS J1054+5832 reported here highlights the effectiveness and potential of the ADFOSC and the 3.6-m DOT for carrying out high sensitivity observations of faint celestial sources in the optical bands. We plan to carry out deep optical observations of other potential HzRGs using the 3.6-m DOT. 

\section*{Acknowledgements}

We thank the staff of ARIES and DOT that made these observations possible. The supports received from Department of Science and Technology, Government of India, Governing council of ARIES, and project management members of the DOT project are duly acknowledged. AO acknowledges academic, technical, and administrative supports received from the staff of ARIES, national and international organizations and industries to design and develop ADFOSC in India. The $4096\times4096$ CCD camera was developed in ARIES, Nainital in technical collaboration with Herzberg Institute of Astrophysics, Canada. 

\vspace{-1em}




\begin{theunbibliography}{} 
\vspace{-1.5em}\bibitem{latexcompanion} 
Bheemireddy K. R. {\em et al.}, 2016, SPIE, 9906, 990644
\bibitem{latexcompanion} 
Best P. N., Longair M. S., R\"ottgering H. J. A., 1998a, MNRAS, 295, 549
\bibitem{latexcompanion} 
Brammer G. B., van Dokkum P. G., Coppi P., 2008, ApJ, 686, 1503
\bibitem{latexcompanion} 
Carilli C. L., Harris D. E., Pentericci L., R\"ottgering H. J. A., Miley G. K., Kurk J. D., van Breugel W., 2002a, ApJ, 567, 781
\bibitem{latexcompanion} 
Chambers K. C. {\em et al.}, 2016, preprint (arXiv:1612.05560)
\bibitem{latexcompanion} 
De Breuck C., van Breugel W., R\"ottgering H. J. A., Miley G., 2000, A\&AS, 143, 303
\bibitem{latexcompanion} 
De Breuck C. {\em et al.}, 2002, AJ, 123, 637
\bibitem{latexcompanion} 
Dunlop J. S., Peacock J. A., 1990, MNRAS, 247, 19
\bibitem{latexcompanion} 
Dye S. {\em et al.}, 2006, MNRAS, 372, 1227
\bibitem{latexcompanion}
Franx M. {\em et al.}, 2003, ApJ, 587, 79
\bibitem{latexcompanion} 
Intema H. T., Jagannathan P., Mooley K. P., Frail D. A., 2017, A\&A, 598, A78
\bibitem{latexcompanion}
Jarvis M. J., Rawlings S., Eales S., Blundell K. M., Bunker A. J., Croft S., McLure R. J., Willott C. J., 2001a, MNRAS, 326, 1585
\bibitem{latexcompanion}
Jarvis M. J., Rawlings S., Willott C. J., Blundell K. M., Eales S., Lacy M., 2001b, MNRAS, 327, 907
\bibitem{latexcompanion}
Ker L. M., Best P. N., Rigby E. E., R\"ottgering H. J. A., Gendre M. A., 2012, MNRAS, 420, 2644
\bibitem{latexcompanion} 
Kumar, B. {\em et al.}, 2018, BSRSL, 87, 29
\bibitem{latexcompanion} 
Lilly S. J., Longair M. S., 1984, MNRAS, 211, 833
\bibitem{latexcompanion}
Mahony E. K. {\em et al.}, 2016, MNRAS, 463, 2997
\bibitem{latexcompanion} 
Miley G., De Breuck C., 2008, A\&AR, 15, 67
\bibitem{latexcompanion} 
Ninane, N., Flebus C., Kumar, B., 2012, SPIE, 2012, 8444, 84441V
\bibitem{latexcompanion} 
Omar A. {\em et al.}, 2012, SPIE, 8446, 844614
\bibitem{latexcompanion} 
Omar, A., Kumar, B., Gopinathan, M., Sagar, R., 2017, CSci, 113, 682.
\bibitem{latexcompanion} 
Omar, A. {\em et al.}, 2019, CSci (Accepted)
\bibitem{latexcompanion}
Orsi A. A., Fanidakis N., Lacey C. G., Baugh C. M., 2016, MNRAS, 456, 3827
\bibitem{latexcompanion}
Pandey S. B. {\em et al.} 2018, BSRSL, 87, 42
\bibitem{latexcompanion} 
Overzier R. A. {\em et al.}, 2009, ApJ, 704, 548
\bibitem{latexcompanion} 
Pentericci L. {\em et al.}, 2000, A\&A, 361, L25
\bibitem{latexcompanion} 
Ree C. H. {\em et al.}, 2011, ApJ, 744, L10
\bibitem{latexcompanion} 
Rigby E. E., Argyle J., Best P. N., Rosario D., R\"ottgering H. J. A., 2015, A\&A, 581, A96
\bibitem{latexcompanion} 
R\"ottgering H. J. A., Lacy M., Miley G. K., Chambers K. C., Saunders R., 1994, A\&AS, 108
\bibitem{latexcompanion}
Rocca-Volmerange B., Le Borgne D., De Breuck C., Fioc M., Moy E., 2004, A\&A, 415, 931
\bibitem{latexcompanion} 
R\"ottgering H., Daddi E., Overzier R., Wilman R., 2003, New Astron. Rev., 47, 309
\bibitem{latexcompanion} 
Saxena A. {\em et al.}, 2018a, MNRAS, 475, 5041
\bibitem{latexcompanion} 
Saxena A. {\em et al.}, 2018b, MNRAS, 480, 2733
\bibitem{latexcompanion} 
Schmidt S. J. {\em et al.}, 2006, ApJ, 649, 63
\bibitem{latexcompanion} 
Steidel C. C., Giavalisco M., Pettini M., Dickinson M.,  Adelberger K. L., 1996, ApJ, 462, L17
\bibitem{latexcompanion} 
van Breugel W. {\em et al.}, 1999, ApJ, 518, L61
\bibitem{latexcompanion} 
Venemans, B. P., Kurk, J. D., Miley, G.K., R\"ottgering, H. J. A., 2003, New Astron. Rev., 47, 353 
\bibitem{latexcompanion} 
White R. L., Becker R. H., Helfand D. J., Gregg M. D., 1997, ApJ, 475, 479
\bibitem{latexcompanion} 
Willott C. J., Rawlings S., Blundell K. M., Lacy M., Eales S. A., 2001, MNRAS, 322, 536
\bibitem{latexcompanion}
Willott C. J., Rawlings S., Jarvis M. J., Blundell K. M., 2003, MNRAS, 339, 173

\end{theunbibliography}
\end{document}